\documentclass[%
 reprint,
superscriptaddress,
 amsmath,amssymb,
 aps,
]{revtex4-2}

\usepackage{graphicx}% Include figure files
\usepackage{dcolumn}% Align table columns on decimal point
\usepackage{bm}% bold math
\usepackage{xcolor}
\usepackage{braket}
\usepackage[utf8]{inputenc}
\usepackage[T1]{fontenc}
\usepackage{etoolbox}

\begin{document}

\preprint{AIP/123-QED}

\title[Magnetically induced transitions for EIT]{Formation of strongly shifted EIT resonances using "forbidden" transitions of Cesium}

%\textcolor{blue}{Application of Cs atomic $F_g = 3 \rightarrow F_e = 5$ transitions in strong magnetic field in electromagnetically induced transparency process}}%SUGGESTION

%Formation of EIT resonances with Cesium forbidden transitions

\author{Armen Sargsyan}
\affiliation{Institute for Physical Research, NAS of Armenia, Ashtarak-2, 0203 Armenia}
\author{Ara Tonoyan}
\affiliation{Institute for Physical Research, NAS of Armenia, Ashtarak-2, 0203 Armenia}
\author{Rodolphe Momier}
\email{rodolphe.momier@u-bourgogne.fr}
\affiliation{Institute for Physical Research, NAS of Armenia, Ashtarak-2, 0203 Armenia}
\affiliation{Laboratoire Interdisciplinaire Carnot de Bourgogne, UMR CNRS 6303, Université Bourgogne Franche-Comté, 21000 Dijon, France}
\author{Claude Leroy}
\affiliation{Laboratoire Interdisciplinaire Carnot de Bourgogne, UMR CNRS 6303, Université Bourgogne Franche-Comté, 21000 Dijon, France}
\author{David Sarkisyan}
\affiliation{Institute for Physical Research, NAS of Armenia, Ashtarak-2, 0203 Armenia}

\date{\today}% It is always \today, today,

\begin{abstract}
Atomic transitions satisfying $F_e - F_g = \Delta F = \pm 2$ (where $F_e$ stands for excited and $F_g$ stands for ground state) of alkali atoms have zero probability in zero magnetic field (they are so-called "forbidden" transitions) but experience a large probabilty increase in an external magnetic field. These transitions are called magnetically induced (MI) transitions. In this paper, we use for the first time the $\sigma^+$ ($\Delta m_F~=~+1$) MI transitions $F_g = 3 \rightarrow F_e = 5$ of {Cesium} as probe radiation to form EIT resonances in strong magnetic fields (1 - 3 kG) while the coupling radiation frequency is resonant with $F_g=4\rightarrow F_e=5$ $\sigma^+$ transitions. The experiment is performed using a nanometric-thin cell filled with Cs vapor and a strong permanent magnet. The thickness of the vapor column is 852 nm, corresponding to the Cs $D_2$ line transition wavelength.
 Due to the large frequency shift slope of the MI transitions ($\sim$ 4 MHz/G), it is possible to form contrasted and strongly frequency-shifted EIT resonances. Particularly, a strong 12 GHz frequency shift is observed when applying an external magnetic field of $\sim$ 3 kG. Preliminary calculations performed considering Doppler-broadened three level systems in a nanocell are in reasonable agreement with the experimental measurements. 

\end{abstract}

\maketitle

\section{\label{sec:intro}Introduction}

\bigbreak

Optical processes occurring in Rubidium, Cesium, Potassium and Sodium vapors confined in optical cells have important applications such as optical atomic clocks, optical atomic magnetometers, atomic gyroscopes, markers of atomic transition frequencies, as described for example in \cite{kitching_chip-scale_2018,vanier_atomic_2005,fleischhauer_electromagnetically_2005,meschede_optics_2008,simons_electromagnetically_2018,abdel_hafiz_protocol_2020}. Therefore, the study of the peculiarities of atomic transitions (in particular Zeeman transitions in an external magnetic field) of alkali atoms is of utmost importance.
%finding new information about peculiarities of atomic transitions of alkali metal atoms, in particular in magnetic fields, is important. 
It is well known that the application of a strong magnetic field can significantly change the probabilities (intensities) of the Zeeman transitions, as shown in \cite{tremblay_absorption_1990,sargsyan_giant_2014,sargsyan_novel_2008,scotto_four-level_2015,tonoyan_circular_2018,sargsyan_circular_2021,pizzey_njp_2022}. 
High interest has recently been focused on atomic transitions between ground and excited levels that satisfy the condition $F_e - F_g = \Delta F = \pm 2$ (these transitions are so-called forbidden by the selection rules, thus their probability is zero when no external magnetic field is applied). However, the probabilities of these transitions in a magnetic field increase significantly. For this reason, we refer to these transitions as Magnetically Induced (MI) transitions \cite{sargsyan_giant_2014,tonoyan_circular_2018,sargsyan_circular_2021}.

This giant increase in the probabilities of the MI transitions is due to the “mixing” of magnetic sublevels $\ket{F,m_F}$ of the ground ($F_g$) or excited ($F_e$) levels with sublevels having the same magnetic quantum number $m_F$. This mixing is the strongest for $D_2$ lines of alkali atoms, as up to four states $\ket{F_e,0}$ can experience mixing, thus resulting in a 4$\times$4 block in the magnetic Hamiltonian, as described in \cite{tremblay_absorption_1990,sargsyan_giant_2014,tonoyan_circular_2018,sargsyan_circular_2021}. 

Magnetically-induced transitions are of great interest because, over a wide range of magnetic field, their probabilities can be much higher than the probabilities of usual ("allowed", satisfying the selection rule on $F$) transitions. It is important to note that the slope of the frequency shifts (obtained by diagonalizing the magnetic Hamiltonian \cite{tremblay_absorption_1990}) as a function of the magnetic field $B$ in strong magnetic fields can reach up to around $4$ MHz/G, which is 3 times larger than in the case of ordinary transitions. Thus, the frequency shift of MI transitions in strong magnetic fields can reach several tens of GHz, which can be useful for working in higher frequency ranges, for example for the frequency stabilisation of lasers on strongly shifted frequencies \cite{sargsyan_saturated-absorption_2014,mathew_simultaneous_2018}.

In \cite{tonoyan_circular_2018,sargsyan_circular_2021}, we established the following rule for the probabilities of MI transitions: the probabilities and number of MI transitions with $\Delta F = +2$ are maximal for $\sigma^+$ radiation, whereas the probabilities and number of MI transitions with $\Delta F = -2$ are maximal for $\sigma^-$ radiation. The difference between the intensities of MI transitions for the $\sigma^+$ and $\sigma^-$-polarized radiation beams can reach several orders of magnitude.

It has been recently demonstrated that electromagnetically-induced transparency (EIT) resonances can be formed using $\Lambda$-system made of $\Delta F = +2$ MI transitions only if both probe and coupling beam are $\sigma^+$-polarized. This statement was experimentally and theoretically verified for $^{87}$Rb (MI transitions $F_g=1\rightarrow F_e=3$) and $^{85}$Rb (MI transitions $F_g=2\rightarrow F_e=4$) \cite{sargsyan_dark_2019,sargsyan_application_2021}. However, if the $\Lambda$-system is formed by MI transitions satisfying $\Delta F = -2$, then both probe and coupling radiation must be $\sigma^-$-polarized in order to form EIT resonances. This statement was experimentally and theoretically verified for Cs (MI transitions $F_g=4\rightarrow F_e=2$). This is a direct consequence of magnetically-induced circular dichroism \cite{sargsyan_coherent_2022}. 

In this work, we consider seven $\sigma^+$ MI transitions of Cs ($F_g = 3 \rightarrow F_e =5$, see Fig.~\ref{fig:fig1}). The probabilities of these transitions increase highly in the range 0.3 - 3 kG and we used these transitions to form EIT resonances in strong $B$-fields. A nanometric-thin cell (NC) filled with Cs vapor (thickness $L \approx 850$ nm, approximately the resonant wavelength of Cs $D_2$ line \cite{SteckCs}) has been used. The advantages of using thin cells, including strong reduction of Doppler broadening, are noted in \cite{sargsyan_circular_2021,sargsyan_application_2021,sargsyan_approach_2019}.

\subsection{Probabilities and frequency shifts of the MI transitions of Cs $D_2$ line}%The Probabilities and Frequency shifts of MI Transitions in the Cs atom, $D_2$ line.}

\begin{figure}
	\includegraphics{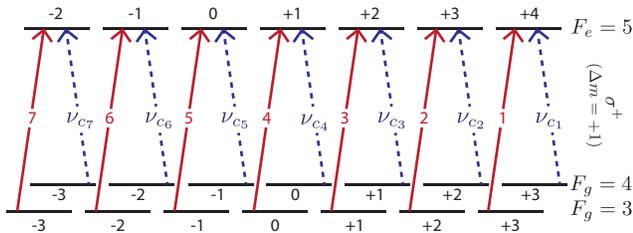}% Here is how to import EPS art
	\caption{\label{fig:fig1} 
	Scheme of Cs $D_2$ line $\sigma^+$ transitions between $F_g = 3,4$ and $F_e = 5$. The probe frequency $\nu_p$ is scanned across the MI transitions labelled 1-7 ($F_g = 3 \rightarrow F_e = 5$). The coupling frequencies $\nu_{c_n}$ are resonant with $F_g = 4 \rightarrow F_e = 5$ transitions, forming seven $\Lambda$-systems. Only the states involved in the process under consideration are shown. Note that $\ket{F,m_F}$ is just a notation for visualization, as the atomic states are better described in the uncoupled basis $\ket{J,m_J,I,m_I}$ in high magnetic fields.}
\end{figure}

\begin{figure*}
	\includegraphics[width=\textwidth]{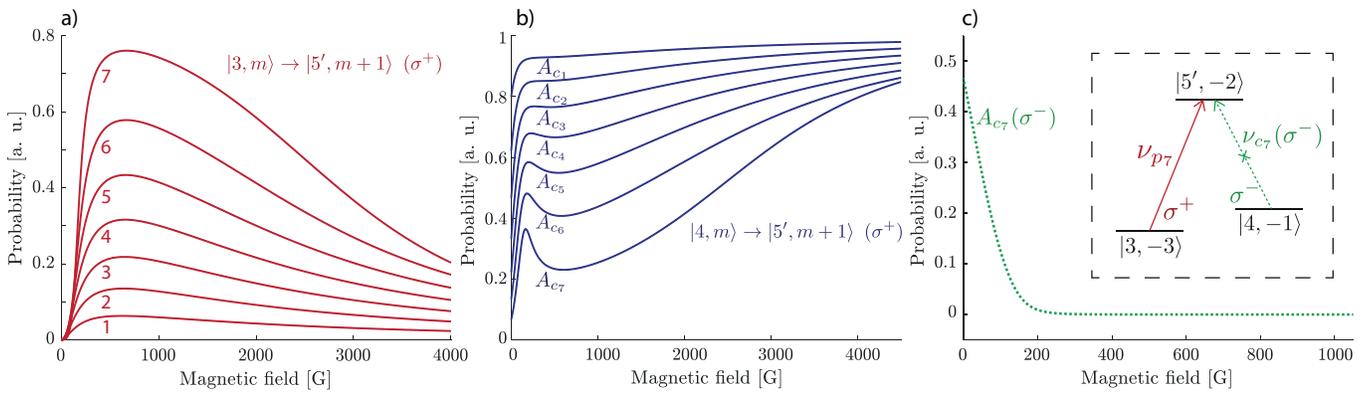}% Here is how to import EPS art
	\caption{\label{fig:fig2} Magnetic field dependence of the Zeeman transition intensities of the $D_2$ line of Cs. a) $F_g = 3 \rightarrow F_e = 5$ $\sigma^+$ MI transitions. b) $F_g = 4 \rightarrow F_e = 5$ $\sigma^+$ transitions. c) Transition $\ket{4,-1}\rightarrow\ket{5,-2}$ $(\sigma^-)$. This transition forms a $\Lambda$-system with transition $7$ as shown in panel a) and in the inset (see Fig.~\ref{fig:fig1}). Its probability tends to 0 as the magnetic field increases, thus forming EIT resonances at high magnetic fields requires both probe and coupling beams to be $\sigma^+$-polarized.}%RODOLPHE 12 OCT}}

\end{figure*}

The curves in Fig.~\ref{fig:fig2} were calculated using a known theoretical model depicting the changes of transition probabilities as a function of the external magnetic field. The block-diagonal (each block corresponding to a given value of the magnetic quantum number) magnetic Hamiltonian is built for each value of the magnetic field and then diagonalized in order to calculate the probability coefficients. This model was presented in a number of papers, e.g. \cite{tremblay_absorption_1990,tonoyan_circular_2018,pizzey_njp_2022}. 

The evolution of the probabilities of MI transitions (labelled 1 to 7, see Fig.~\ref{fig:fig1}) with respect to the magnetic field $B$ is shown in Fig.~\ref{fig:fig2}a). Note that in the range 0.3 - 2 kG the probabilities of the MI transitions labeled 5, 6 and 7 are the strongest among all transitions occurring from $F_g=3$ \cite{sargsyan_giant_2014,sargsyan_circular_2021}. The frequency shift slope of the MI transitions, obtained through the eigenvalues of the Hamiltonian, is quite large ($\sim$ 4 MHz/G) while for usual transitions the slope is 3 times smaller. Despite the fact that the probabilities of the MI transitions decrease as $B$ increases, they can still be recorded easily at 7 kG. As noted below, this is due to the fact that these transitions are formed far on the high-frequency wing where there are no intersections with other transitions (spectra are presented for Na in \cite{momier_sodium_2021}, but Cs behaves almost identically).

The evolution of the probabilities of the corresponding seven coupling transitions $F_g=4 \rightarrow F_g=5$ ($A_{c_1}$ to $A_{c_7}$) that are used to form seven $\Lambda$-systems (see Fig.~\ref{fig:fig1}) with respect to the magnetic field are shown in Fig.~\ref{fig:fig2}b). In the case of $\sigma^-$ polarization, the probability of the strongest $F_g = 4 \rightarrow F_e = 5$ $\sigma^-$ transition already tends to zero for $B > 300$ G, as shown in Fig.~\ref{fig:fig2}c). Thus, both the probe and the coupling beams must be $\sigma^+$-polarized in order to form EIT resonances.

\subsection{\label{sec:EIT}Qualitative description of the EIT process}

For a qualitative description of the EIT process, we present a formula from \cite{fleischhauer_electromagnetically_2005,gea-banacloche_electromagnetically_1995}. The ratio of absorption at the probe radiation frequency $\nu_p$ at which EIT resonance is observed (in the presence of $\nu_c$ radiation) to absorption (when there is no coupling radiation), assuming low radiation intensity $\nu_p$ and zero frequency detuning of the coupling radiation, is described by the expression:

\begin{equation}
 	\frac{\alpha(\Omega_c)}{\alpha(0)} = \frac{K}{1+\Omega_c^2/4\Gamma_{21}\gamma_N}\, ,\label{eq:depth}
\end{equation}

where $K$ is a constant including the Doppler width, $\gamma_N$ is the natural width of the level ($\gamma_N/2\pi~\simeq~5.2$ MHz for the $6^2P_{3/2}$ level of the Cs atom), $\Omega_c$ is the Rabi frequency for the coupling radiation and $\Gamma_{21}$ is the dephasing rate of the coherence between the two ground states of the $\Lambda$-system, which is caused in particular by collisions of atoms with the windows of the nanocell. The case $\alpha(\Omega_c) = 0$ corresponds to complete transparency (the contrast of the EIT resonance reaches 100\%) and a large amplitude of the EIT resonance, which decreases with an increase in $\Gamma_{21}$. The spectral width of the EIT resonance can be described by the simple expression \cite{fleischhauer_electromagnetically_2005}:

\begin{equation}
	\gamma_{_\text{EIT}} \simeq 2\Gamma_{21} + \Omega_c^2/\gamma_N\, . \label{eq:width}
\end{equation}

It follows from formula (\ref{eq:depth}) that in order to obtain small value of $\alpha(\Omega_c)$ (which means high electromagnetically induced transparency of the medium), it is necessary to increase $\Omega_c$, however, an increase in $\Omega_c$ leads to an increase in the spectral width of the EIT resonance. Therefore, it is necessary to find a compromise for $\Omega_c$. Estimates can be obtained from $\Omega_c/2\pi = a \gamma_N ( I / 8)^{1/2}$ where $I$ is the laser intensity in mW/cm$^2$, $\gamma_N \sim 5$ MHz, and $a$ is a fit parameter (for our case $a$ is of $\sim$ 0.5) \cite{grove_two-photon_1995} and $\Omega_c \sim 15$~MHz.

\section{\label{sec:exp}Experiment}
\subsection{\label{sec:setup}Experimental setup}
\begin{figure}
	\includegraphics[scale=1.2]{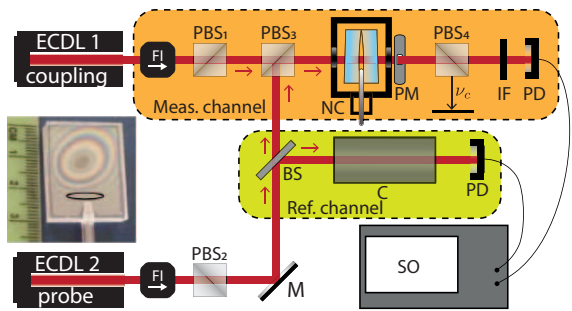}% Here is how to import EPS art
	\caption{\label{fig:fig3} Scheme of the experimental setup. ECDL: CW narrow-band external-cavity diode lasers with $\lambda = 852$ nm (resonant with Cs $D_2$ line). FI: Faraday insulators. PBS$_i$: polarizing beam splitters. BS: beam splitter. IF: interference filter. C: saturated absorption spectroscopy unit for frequency reference. NC: nanocell placed in oven. PM: permanent magnet. PD: photodiodes. SO: 4-channel digital oscilloscope.}% The reference channel corresponds to a DAVLL frequency-locking scheme, of which the representation is simplified for the sake of clarity. The picture represents the cell used in the experiment, and the oval corresponds to the area where $L \approx \lambda$.}%RODOLPHE 12 OCT}}
	%Schematic of the experimental setup: ECDL1,2 cw narrow-band external-cavity diode lasers with $\lambda$ =852 nm; (1) Faraday insulators; (PBS1, PBS2, PBS3, PBS4) polarizing beam splitters; (2) T-shaped Cs NC, (3) Cs NC in a furnace,(4) photo-receivers; (5) unit for frequency reference SAS formation ; (6) 4-chanel digital oscilloscope; (IF) interference filter, PM-permanent magnet.}
\end{figure}

\begin{figure}
	\includegraphics[scale=0.8]{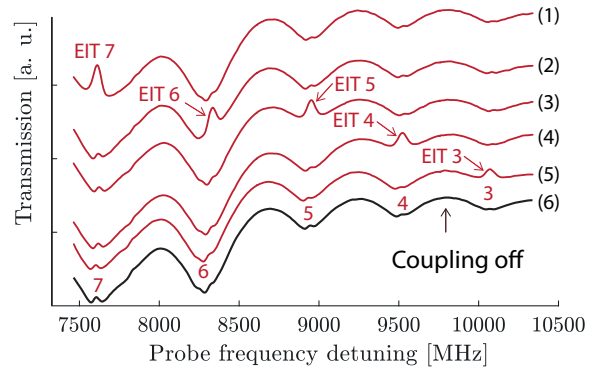}% Here is how to import EPS art
	\caption{\label{fig:fig4} 
	Probe transmission spectra of the Cs nanocell ($L = \lambda = 852$ nm). Spectra labelled 1 to 5 show five EIT resonances, labelled EIT 3 to EIT 7, while the probe frequency is scanned across transitions 3 to 7 (see Fig.~\ref{fig:fig1}). The coupling and probe powers are respectively 10 and 0.05 mW and the external longitudinal magnetic field is $B = 1400$ G. Spectrum n° 6 corresponds to the case where coupling is off. Small VSOP peaks are visible on each atomic resonance. Zero frequency corresponds to the transition frequency of Cs $D_2$ line.}%RODOLPHE 12 OCT}}
	
	% (ORIGINAL) Cs NC, L=852nm, lines (1)-(5) show the
	%	transmission spectra of the probe radiation,. which contain
	%	EIT7, 6, 5, 4, 3 resonances, the MI transitions with numbers
	%	3-7 are involved, respectively, a longitudinal magnetic field
	%	B=1400 G was applied. The coupling and the probe powers
	%	for the $\sigma^+$ polarizations are 10 and 0.05 mW, respectively. 4
	%	Line (6) shows only the probe spectrum and there are small VSOP peaks that are located at the atomic transitions. \textcolor{red}{Zero frequency corresponds to the fine transition $6S_{1/2}$ to $6P_{3/2}$ for Cs $D_2$ line.}}
\end{figure}

The layout of the experimental setup is shown on Fig.~\ref{fig:fig3}. Two extended cavity diode lasers are tuned in the vicinity of the Cs $D_2$ line, with a wavelength $\lambda \simeq 852$ nm. The $\Lambda$-systems shown in Fig.~\ref{fig:fig1} are formed by scanning the frequency $\nu_p$ of a VitaWave laser ($\delta\nu_p\sim 1$ MHz) \cite{vassiliev_compact_2006} in the vicinity of the MI transitions $F_g=3 \rightarrow F_e=5$, while keeping the frequency $\nu_c$ from a MOGLabs “cateye” laser ($\delta\nu_p \simeq 0.1$ MHz) on resonance with one of the $4 \rightarrow 5$ transitions. A fraction about 10\% of the coupling radiation power was sent to a frequency stabilization unit based on the DAVLL method \cite{yashchuk_laser_2000}. Probe radiation has vertical polarization, while the coupling radiation has horizontal polarization. In the case of a longitudinal B-field, linearly polarized laser radiation can be considered as consisting of $\sigma^+$ and $\sigma^-$ radiations. The use of mutually perpendicular polarizations allows by using PBS4 to direct only probe radiation to the photo-receiver, while cutting off the coupling radiation. As noted above, in the case of MI transitions with $\Delta F=+2$ for the formation of the EIT resonance, both probe and coupling radiations must have $\sigma^+$ polarization. A photograph of the Cs nanocell is shown in Fig.~\ref{fig:fig3}. Interference fringes are formed by the reflection of light on the inner surfaces of windows (made of sapphire). The region corresponding to a thickness $L \approx \lambda \sim 850$ nm is outlined by an oval. The design of the Cs-filled NC used in our experiments is similar to that of extremely thin cell described in \cite{keaveney_maximal_2012}. Earlier it was demonstrated in \cite{sargsyan_dark_2019,sargsyan_application_2021,sargsyan_high-contrast_2015} that the use of a nanocell (NC) with thickness $L=\lambda$ makes it easy to record contrasted EIT resonances, which is due to the low absorption of the NC, while the disadvantage is broadening of the EIT resonance caused by frequent inelastic collisions of atoms with the windows of the NC. Studies of the EIT resonances were done using a strong neodymium–iron–boron alloy ring-shaped permanent magnet (PM). Due to the small thickness of the vapor column, the high-gradient field produced by magne can be considered uniform across the interaction region. The PM was placed after the rear window of the NC, with the axis aligned along the probe beam propagation direction. The magnetic field in the NC was simply varied by longitudinal displacement of the PM, calibrated using a Teslameter HT201 magnetometer.

\subsection{Experimental results: using MI transitions to form EIT resonances}

Curves 1 to 5 in Fig.~\ref{fig:fig4} show the experimental transmission spectra of the probe radiation which contain the resonances EIT 3 to EIT 7 (numbers 3-7 means that MI transitions with numbers 3-7 are involved, respectively) in a longitudinal magnetic field $B = 1400$ G. The NC thickness is $L=\lambda=852$ nm and the temperature of the reservoir is 100 $^\circ$C (to prevent Cs vapor condensation on the windows, the temperature of the windows is slightly higher). The coupling and the probe powers are 20 mW and 0.1 mW, respectively. Note that since only $\sigma^+$ radiations participate to the formation of the EIT resonances (see Fig.~\ref{fig:fig1}), only half of the power of these radiations must be considered, meaning 10 mW and 50 $\mu$W, respectively. Curve n° 6 is a probe spectrum when the coupling is blocked. Since the cell thickness is $L=\lambda$, small peaks formed by velocity selective optical pumping (VSOP) resonances are located exactly at the atomic transitions frequencies, as described in \cite{sargsyan_novel_2008}. 

The amplitude of the EIT resonance is a factor $\sim$10 larger than the amplitude of the VSOP resonance, whereas the spectral width of the EIT resonance is a factor of 1.5 smaller, which is characteristic of the coherent EIT process \cite{sargsyan_application_2021}. Note that the contrast of the EIT resonance defined as the ratio of the EIT resonance amplitude divided by the peak absorption of the Cs vapor when the coupling is blocked reached 40-50 \% which is typical when a nanocell is used \cite{sargsyan_high-contrast_2015}.

In Fig.~\ref{fig:fig6}, curves 1 to 4 are probe transmission spectra which contain EIT 6, EIT 5, EIT 4 and EIT 3 resonances for $B = 1770$ G. Curve n° 5 shows only the probe spectrum when the coupling is blocked. In Fig.~\ref{fig:fig7}, lines 1 to 3 show the probe transmission spectra which contain EIT 6, EIT 4 and EIT 3 resonances for $B = 2880$ G. Line n° 4 shows only the probe spectrum when the coupling is blocked. The inset
shows the profile of EIT 6 resonance fitted with a Gaussian profile with a FWHM of $\sim$ 35 MHz. There is also a small VSOP resonance which is formed when the coupling is blocked. The typical FWHM of VSOP resonances is 40-50 MHz. 

\begin{figure}[h]
	\includegraphics[scale=2.2]{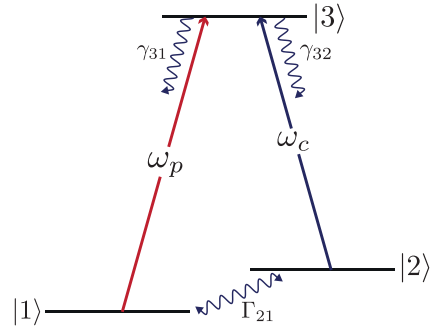}% Here is how to import EPS art
	\caption{\label{fig:fig5} Scheme of the three-level $\Lambda$-system used in the calculations. The total decay rate $\Gamma_{33}$ of state $\ket{3}$ is $1/2(\gamma_{31} + \gamma_{32})$ \cite{Shore}. The dephasing rate of coherence between the ground states is $\Gamma_{21} = (2\pi t)^{-1}$ where $t$ is the time of flight of the atoms through the cell (at the most probable velocity $u = \sqrt{2k_BT/M}$ where $T$ is the vapor temperature and $M$ the atomic mass).}%RODOLPHE 12 OCT}}
	%Schematic of the experimental setup: ECDL1,2 cw narrow-band external-cavity diode lasers with $\lambda$ =852 nm; (1) Faraday insulators; (PBS1, PBS2, PBS3, PBS4) polarizing beam splitters; (2) T-shaped Cs NC, (3) Cs NC in a furnace,(4) photo-receivers; (5) unit for frequency reference SAS formation ; (6) 4-chanel digital oscilloscope; (IF) interference filter, PM-permanent magnet.}
\end{figure}

\begin{figure}
	\includegraphics[scale=0.8]{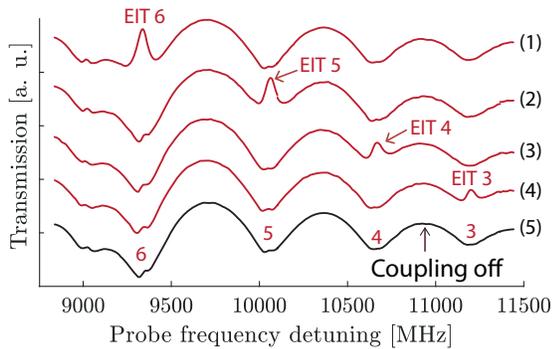}% Here is how to import EPS art
	\caption{\label{fig:fig6} 
		Probe transmission spectra of the Cs nanocell ($L~=~\lambda \approx 850$ nm). Spectra 1 to 4 exhibit four EIT resonances, labelled EIT 3 to EIT 6, while the probe frequency is scanned across transitions 3 to 6. The external longitudinal magnetic field is $B = 1770$ G. Spectrum n° 5 is a probe transmission spectrum when the coupling is off. Small VSOP peaks are visible on each atomic transition. Zero frequency corresponds to the transition frequency of Cs $D_2$ line.} %RODOLPHE 12 OCT}}

	%(ORIGINAL) Cs NC, L=852nm, a) lines (1)- (4) shows the probe experimental transmission spectra which contain the EIT-6, 5, 4 ,3 resonances, B-field is 1770 G, the line (5) shows only the probe spectrum and there are small VSOP peaks that are located at the atomic transitions.
	%EIT 4}
\end{figure}

\begin{figure}
	\includegraphics[scale=0.8]{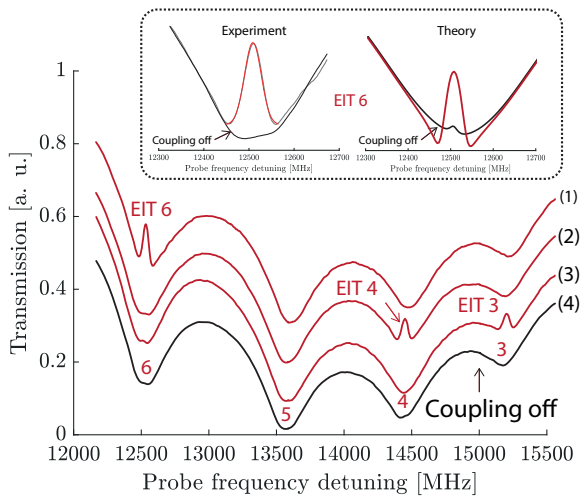}% Here is how to import EPS art
	\caption{\label{fig:fig7}
	Probe transmission spectra of the Cs nanocell ($L = \lambda = 852$ nm). Lines 1 to 3 show four EIT resonances, labelled EIT 4, EIT 5 and EIT 6. The external longitudinal magnetic field is $B = 2880$ G. Line 4 is a probe transmission spectrum when the coupling is off. The left part of the inset is a zoom on EIT 6, fitted with a Gaussian profile (FWHM 35 MHz). The right curves are calculated. Red: coupling on, black: coupling off. Small VSOP peaks are visible on each atomic transitions formed by the probe radiation. Their typical linewidth is 40-50 MHz. Zero frequency corresponds to the transition frequency of Cs $D_2$ line.}%RODOLPHE 12 OCT
	%(ORIGINAL) Lines (1)-(3) show the probe transmission spectra containing EIT-6, 4, 3 resonances, B= 2800 G, the line (4) shows only the probe spectrum and there are small VSOP peaks which are located at the atomic transitions. The inset shows the profile of EIT-6 resonance which is fitted with the Gaussian profile, and full width halves maximum (FWHM) is of $\sim$ 35 MHz. There is also a small VSOP resonance which is formed by the probe radiation, typically VSOPs linewidth FWHM is 40-50 MHz.}
\end{figure}

Preliminary theoretical calculations (shown in the right part of the inset of Fig.~\ref{fig:fig7} were obtained by solving the Liouville equations of motion for an ensemble of three-level $\Lambda$-systems (as presented in Fig.~\ref{fig:fig5}), taking into account the geometry of the nanocell (coherence dephasing rate determined by the time of flight of the atoms), its Fabry-Perot nature (reflections of the fields on the inner surfaces of the cell) and Doppler broadening, following the procedure described in \cite{pashayan_JOSAB_EIT}. The Rabi frequencies of the probe and coupling lasers are respectively $\Omega_c = 1.5\gamma_N$ and $\Omega_p = 0.06\gamma_N$.
Reasonable agreement between theory and experiment regarding the width and depth of the EIT resonance is obtained and the VSOP resonance is seen. Small discrepancies (assymetry of the profile and amplitude of the VSOP resonance) can arise notably from the need of considering neighboring Zeeman sublevels (not shown in Fig.~\ref{fig:fig1}, and therefore more than three levels, to obtain more accurate results.

\begin{figure}
	\includegraphics[scale=0.8]{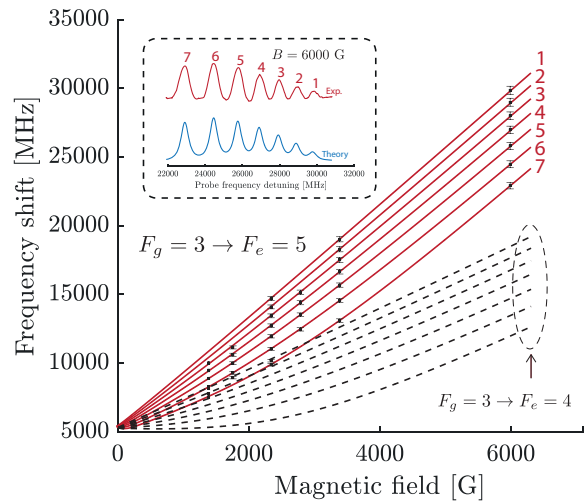}% Here is how to import EPS art
	\caption{\label{fig:fig8} Red solid lines: frequency shift of transitions 1 to 7 (see figure \ref{fig:fig1}) as a function of the magnetic field. The black squares with error bars represent experimental measurements, the inaccuracy is around 1 \%. Black dashed lines: frequency shift of $F_g = 3 \rightarrow F_e = 4$ transitions. For $B > 3$ kG, both groups are well separated in frequency. Inset: theoretical and experimental absorption spectra for $B = 6$ kG, the frequency shift reaches 30 GHz from the Cs $D_2$ line transition frequency.}%RODOLPHE 12 OCT

	%(ORIGINAL) The solid lines indicate the dependences of the calculated frequency shifts for the Cs D2 MI transitions 1-7 and for $F_g = 3 \rightarrow F_e = 4$ transitions vs B-field; the black squares, represent the experimental results. For $B > 3$ kG the group of MI transitions is completely separated in frequency from $F_g = 3 \rightarrow F_e = 4$ transitions, which is important for an applications. The inaccuracy is of $\sim$ 1\%. The upper and lower lines in the inset show experimental and theoretical spectra of the MI transitions absorption for $B = 6$ kG. The frequency shift reached $\sim$ 30 GHz.}
\end{figure}

The amplitude of resonance n° 6 is $\sim$ 50 times greater than that of the VSOP resonance and is spectrally narrower than the latter (this is a manifestation of the coherent EIT process \cite{vanier_atomic_2005,sargsyan_application_2021}). In Fig.~\ref{fig:fig8} the solid lines indicate the calculated dependences of the frequency shifts for transitions 1–7 (Fig.~\ref{fig:fig1}) and $F_g=3\rightarrow F_e=4$ (marked with dotted oval) to the magnetic field $B$. The black squares represent the experimental results. As mentioned earlier, due to the high value of the frequency shift slope for $B>3$ kG, the group of MI transitions 1–7 is completely separated in frequency from $F_g = 3 \rightarrow F_e = 4$ transitions.

The curves in the inset of Fig.~\ref{fig:fig8} show experimental and theoretical spectra (calculated by combining the models presented in \cite{tremblay_absorption_1990} and \cite{dutier_JOSAB}) of the seven MI transitions absorption for $B = 6$ kG when frequency shift reaches $\sim$ 30 GHz. Note that the amplitude of transition 6 is slightly bigger than that of transition 7 (while for $B < 5$ kG the amplitude of transition 7 is bigger, see Fig.~\ref{fig:fig2}a),  because of the “mixing” effect. 
%The MI transitions are still well recorded up to 7 kG and for this case the frequency shift reaches $\sim$ 34 GHz. This is a remarkable feature of the MI $F_g=3\rightarrow F_e=5$ transitions. The MI transitions at such high $B$-fields can still be exploited if the frequency tuning of the diode lasers used is very large (in our case, the linear frequency tuning range was 5-6 GHz).
Note that one of the remarkable features of the $\sigma^+$ MI transitions $3 \rightarrow 5'$ is that they are still well recorded for a magnetic field $B\approx8$ kG. They are located in the high frequency wing of the spectrum presented in Fig.~18 of paper \cite{staerkind_arxiv_2022} and for this case the frequency shift reaches 34 GHz. Using our theoretically calculated curves for MI transitions $3\rightarrow 5'$ we checked the frequency position of these MI transitions and found good agreement with the experimental curves presented in Fig.~18. In paper \cite{staerkind_arxiv_2022} the $3\rightarrow 5'$transitions are not identified. Therefore, it is important to inform scientists working in the field of laser spectroscopy of alkali metal atoms about the MI atomic transitions. The above-mentioned MI transitions can be exploited in such high B-fields as new frequency markers, for using new frequency ranges, as well as for the frequency stabilization of lasers at strongly shifted frequencies from the initial transition in unperturbed atoms \cite{pizzey_njp_2022,sargsyan_saturated-absorption_2014}.

\section{Conclusion}

In this paper, we used for the first time forbidden transitions of Cs ($F_g = 3 \rightarrow F_e = 5$, more precisely $\sigma^+ (\Delta m_F = +1)$ transitions) to create $\Lambda$-system allowing the formation of EIT resonances. This was done in an external magnetic field, as such transitions have zero probability in the absence of magnetic field. A nanometric-thin cell filled with Cs vapor was used, with a thickness corresponding to the resonant wavelength of Cs $D_2$ line ($\approx 850$ nm), and the magnetic field was varied by longitudinal displacement of the permanent magnet along the propagation direction (Fig.~\ref{fig:fig3}). As expected, when the coupling is blocked, small VSOP resonances are formed right at the different transitions' frequencies, while coupling radiation allows for the formation of EIT resonances, spectrally narrower and with a bigger amplitude. 
We formed EIT resonances with 6 out the 7 transitions depicted in Fig.~\ref{fig:fig1}. This was possible up to 3 kG thanks to the big value of the frequency shift, reaching up to 4 MHz/G, therefore leading to EIT resonances shifted 12 GHz apart from the Cs $D_2$ line transition frequency \cite{staerkind_arxiv_2022}. This result is of great interest, as the highly-shifted spectra can serve as frequency references \cite{sargsyan_saturated-absorption_2014,mathew_simultaneous_2018}, especially taking into account that these transitions are still easily recorded up to 8 kG when the frequency shift reaches 35 GHz. 
As for the theoretical description, further investigation is necessary, mainly in order to take into account the effect of neighbouring states, and thus including more levels in the model. The complexity of the manifold and the number of coupled equations make it a challenging and computationally-intensive task. However, reasonable agreement was already achieved by simply considering an ensemble of three-level systems. 
To the best of our knowledge, there are no reports on obtaining EIT resonances in $\Lambda$-systems in such strong fields using usual transitions of alkali atoms. We note that much narrower EIT resonances can be attained by using cm-long cells (to lower the effect of inelastic collisions of atoms with the windows), and by using coherently coupled probe and coupling radiations derived from a single narrow-band laser beam \cite{fleischhauer_electromagnetically_2005}.

\begin{acknowledgments}
This work was supported by the Science Committee of the Republic of Armenia, in the frame of research project n° 21T-1C005, and by the NATO Science for Peace and Security Project under grant G5794. 
\end{acknowledgments}

\section*{Data Availability Statement}
Data underlying the results presented in this paper are not publicly available at this time but may be obtained from the authors upon reasonable request.

\nocite{*}
\bibliography{biblio_oct22}% Produces the bibliography via BibTeX.

\end{document}